\documentclass[twocolumn,showpacs,aps]{revtex4}
\usepackage{amsmath}
\usepackage{dcolumn}
\usepackage{graphicx}
\usepackage{bm}

\begin{document}

\title{Mott insulating phases and quantum phase transitions of interacting
spin-$\frac{3}{2}$ fermionic cold atoms in optical lattices at half filling}
\author{Hong-Hao Tu and Guang-Ming Zhang}
\email{gmzhang@tsinghua.edu.cn}
\affiliation{Department of
Physics, Tsinghua University, Beijing 100084, China}
\author{Lu Yu}
\affiliation{Institute of Physics and Institute of Theoretical
Physics, Chinese Academy of Sciences, Beijing 100080, China}
\date{\today }

\begin{abstract}
We study various Mott insulating phases of interacting spin-$\frac{3}{2}$
fermionic ultracold atoms in two-dimensional square optical lattices at half
filling. Using a generalized one-band Hubbard model with hidden SO(5)
symmetry, we identify two distinct symmetry breaking phases: the degenerate
antiferromagnetic spin-dipole/spin-octupole ordering and spin-quadrupole
ordering, depending on the sign of the spin-dependent interaction. These two
competing orders exhibit very different symmetry properties, low energy
excitations and topological characterizations. Near the SU(4) symmetric
point, a quantum critical state with a $\pi $-flux phase may emerge due to
strong quantum fluctuations, leading to spin algebraic correlations and
gapless excitations.
\end{abstract}


\maketitle

\section{Introduction}

The Mott insulating states are attracting extensive attention in modern
condensed matter physics, which are believed to play a crucial role for
resolving the mysteries in high temperature superconductors, heavy fermion
compounds, colossal magnetoresistance manganites and so on. In the
conventional band theory, the insulators are characterized by an energy gap
between the highest filled and the lowest empty states. On the other hand, a
charge excitation gap can open up with a partially filled energy band due to
the competition between kinetic energy and repulsive force between the
particles. Since the strong correlation effects are difficult to handled, an
efficient theoretical description of the Mott insulating states is still a
challenging task.

Since Greiner \textit{et al.} experimentally succeeded in observing the
superfluid-Mott phase transition of ultracold bosonic atoms in optical
lattices \cite{Bloch-2002}, studying strongly interacting cold atom systems
becomes a new research direction of condensed matter physics. In fact, an
atomic Mott insulator is prepared using a deep optical lattice, where the
particle number fluctuations are strongly suppressed. Such systems are
rather clean and controllable compared to conventional solid state systems.
Besides the single component atoms, the spinor atoms can also be loaded into
the optical traps or lattices, leading to a variety of novel quantum
phenomena.

Depending on the hyperfine spin values, the ultracold atoms are classified
as bosons with integer $F$ and fermions with half odd $F$. Most previous
studies on the Mott insulating states focused on scalar and spinor bosonic
atoms. Various Mott insulating phases are also formed in higher spin
ultracold atoms in optical lattices. For spin-1 bosons, the spin-singlet,
dimerized, and the nematic Mott insulating states were studied \cite%
{Demler-Zhou-2002,Yip-2003,Zhou-2003,Demler-2003,Bernier-Kim-2006}. For
spin-2 bosons, even more complicated Mott insulators can be realized \cite%
{Zhou-2006,Eckert-2006}. Depending on the scattering length and particle
occupation number per site, the ground state can be spin-ordered
ferromagnetic, cyclic and nematic states. When the potential barriers of the
optical lattices are very high, one can even observe the coherent
spin-mixing oscillations of the two isolated spin-2 $^{87}$Rb atoms in one
site \cite{Bloch-2005,HJHuang-2006}. Recently, the Mott insulating phases of
spin-3 $^{52}$Cr atoms have attracted attention motivated by the
experimental progress, where the long-range dipole interaction plays an
important role in determining the exotic ground states \cite{Bernier-2006}.

Actually, the physics of fermionic ultracold atoms in optical lattices is
also interesting. For spin-$\frac{1}{2}$ fermions, the Hubbard and $t-J$
like models can be realized in optical lattices, which may shed light on the
fundamental issues in studying the high-$T_{c}$ superconductors in the doped
cuprates \cite{klein-jaksch}. Furthermore, the high spin fermionic atoms
provide opportunities to study novel basic physics with no counterparts for
electrons in solids. Recently, rare-earth $^{173}$Yb atoms with the lowest
hyperfine manifold $F=\frac{5}{2}$ have been cooled down to quantum
degeneracy regime \cite{Takahashi-2007}, free of spin relaxation. On the
other hand, the $F=\frac{3}{2}$ interacting fermionic system is attracting
theoretical interest, which can be realized with alkali atoms $^{132}$Cs, as
well as alkaline-earth atoms $^{9}$Be, $^{135}$Ba, and $^{137}$Ba. Recently,
Wu \textit{et al.} noticed that the spin-$\frac{3}{2}$ fermions with $s$%
-wave scattering interaction enjoy a hidden SO(5) symmetry, without tuning
any parameter \cite{CWu-2003,CWu-2006}. Then, more interesting issues are
studied, such as the competition between the baryon-like quartet pairing
state and quintet pairing state in one-dimension \cite{CWu-2005}, the
topological generation of quantum entanglement by non-Abelian half quantum
vortices in quintet superfluid phase \cite{CWu-Hu-2005}.

In this paper, we will investigate the various Mott-insulating states and
map out the ground state phase diagram for a half filled spin-$\frac{3}{2}$
ultracold fermionic atom system on a two dimensional square optical lattice.
In addition to the spin-quadrupole ordering state \cite{Tu-2006}, we will
show that a new Mott insulating state with degenerate antiferromagnetic
spin-dipole/spin-octupole ordering can be formed as well. Between these two
distinct spin-ordered phases, an SU(4) $\pi $-flux state may emerge as a
quantum critical phase due to the competing orders. Furthermore, the
spin-quadrupole correlations, as well as the spin-dipole and spin-octupole
correlations display the same power-law behavior. This is in analogy with
the staggered flux state in quantum spin $S=\frac{1}{2}$ Heisenberg
antiferromagnets, in which the competing N\'{e}el order parameter and the
valence-bond solid order parameter exhibit the same power-law correlations.
In fact, the SU(4) $\pi $-flux state and the SU(2) staggered flux state both
belong to a class of algebraic spin liquid states that unify the competing
orders by emergent symmetry at low energies \cite{Hermele-2005}. A variety
of novel algebraic spin liquid states have been studied in the underdoped
region of cuprates \cite{Rantner-2001}, toy spin models \cite%
{Moessner-2003,Wen-2003,Hermele-2003,Xu-2006} and strongly correlated
\textit{p}-orbital band bosons in optical lattices \cite{Xu-Fisher-2006}.

The paper is organized as follows. In Sec. II, the general spin-$\frac{3}{2}$
Hubbard model will be introduced, which describes the spin-$\frac{3}{2}$
interacting fermionic atoms in optical lattices. In Sec. III, we will
briefly review the Mott insulating phase with a staggered spin-quadrupole
order. In Sec. IV, the Mott insulating phase with degenerate
antiferromagnetic spin-dipole/spin-octupole order will be studied within the
functional integral approach. The corresponding symmetry breaking pattern
and the low energy collective excitation modes are discussed. Sec. V is
devoted to an analysis of the SU(4) $\pi $-flux state as a quantum critical
state. The phase diagram is obtained based on an effective model at low
energies. Finally, a brief discussion is presented in Sec. VI.

\section{Generalized one-band Hubbard model}

An optical lattice is a periodic potential $V(\mathbf{r})=%
\sum_{r=x,y,z}V_{r}\sin ^{2}(\mathbf{k\cdot r})$ created by standing-wave
laser beams, where $|\mathbf{k}|=2\pi /\lambda $ is the wave vector of the
laser and $\lambda /2$ is the lattice spacing. When the potential is tuned
to $V_{z}\gg V_{x}=V_{y}$, the atoms are confined to the lowest Bloch band
of a two-dimensional $x$-$y$ plane square lattice at low temperatures. The
system of spin-$\frac{3}{2}$ ultracold interacting fermionic atoms is well
described by a generalized one-band Hubbard model%
\begin{eqnarray}
H &=&-t\sum_{\langle ij\rangle }(\psi _{i}^{\dag }\psi _{j}+\text{H.c.})+%
\frac{c_{0}}{2}\sum_{i}(N_{i}-2)^{2}  \notag \\
&&+\frac{c_{2}}{2}\sum_{i}\mathbf{S}_{i}^{2}-\mu \sum_{i}N_{i},
\end{eqnarray}%
where $\psi $ is a four-component spinor defined by $\psi _{i}=\left( \psi
_{i,3/2},\psi _{i,1/2},\psi _{i,-1/2},\psi _{i,-3/2}\right) ^{T}$ and the
chemical potential $\mu =0$ is fixed at half filling on a bipartite lattice.
Within the harmonic approximation, the parameters for the two-dimensional
square optical lattice are given by%
\begin{eqnarray}
t &=&\frac{2}{\sqrt{\pi }}E_{R}\xi ^{3}\exp (-2\xi ^{2}),  \notag \\
c_{0} &\sim &\frac{5a_{2}-a_{0}}{\lambda }E_{R}\xi ^{3},  \notag \\
c_{2} &\sim &\frac{a_{2}-a_{0}}{\lambda }E_{R}\xi ^{3},
\end{eqnarray}%
where $E_{R}=\hbar k^{2}/2m$ is the recoil energy, $\xi =(V_{x}/E_{R})^{1/4}$
and $a_{S}$ is the $s$-wave scattering length in the total spin-$S$ channel.
Due to the Pauli's exclusion principle, the $s$-wave interactions of
identical spin-$\frac{3}{2}$ fermions in total spin $S=1$, $3$ channels are
forbidden. The particle number and spin operators at lattice site $i$ are
defined by%
\begin{equation}
N_{i}=\sum_{\alpha }\psi _{i\alpha }^{\dag }\psi _{i\alpha },\text{ }\mathbf{%
S}_{i}=\sum_{\alpha \beta }\psi _{i\alpha }^{\dag }\mathbf{S}_{\alpha \beta
}\psi _{i\beta },
\end{equation}%
where $\mathbf{S}$ are the spin-$\frac{3}{2}$ matrices. In addition, we have
to consider the higher-order products of the spin matrices. The
spin-quadrupole matrices are defined by
\begin{eqnarray}
\frac{1}{\sqrt{3}}\left( S^{x}S^{y}+S^{y}S^{x}\right) &=&\Gamma ^{1},\text{ }%
\frac{1}{\sqrt{3}}\left( S^{z}S^{x}+S^{x}S^{z}\right) =\Gamma ^{2},  \notag
\\
\frac{1}{\sqrt{3}}\left( S^{z}S^{y}+S^{y}S^{z}\right) &=&\Gamma ^{3},\text{ }%
\left( S^{z}\right) ^{2}-\frac{5}{4}=\Gamma ^{4},  \notag \\
\frac{1}{\sqrt{3}}\left[ (S^{x})^{2}-(S^{y})^{2}\right] &=&\Gamma ^{5},
\end{eqnarray}%
where%
\begin{equation*}
\Gamma ^{1}=\left(
\begin{array}{cc}
0 & -iI \\
iI & 0%
\end{array}%
\right) ,\Gamma ^{2,3,4}=\left(
\begin{array}{cc}
\vec{\sigma} & 0 \\
0 & -\vec{\sigma}%
\end{array}%
\right) ,\Gamma ^{5}=\left(
\begin{array}{cc}
0 & I \\
I & 0%
\end{array}%
\right) ,
\end{equation*}%
are the five Dirac Gamma matrices satisfying the Clifford algebra $\{\Gamma
^{a},\Gamma ^{b}\}=2\delta _{ab}$, $I$ is a $2\times 2$ unit matrix and $%
\vec{\sigma}$ are Pauli matrices. In terms of $\Gamma $ matrices, we have
also tensor operators defined by $\Gamma ^{ab}=-\frac{i}{2}[\Gamma
^{a},\Gamma ^{b}]$. Thus, the spin-$\frac{3}{2}$ (dipole) matrices are
expressed as linear combinations of $\Gamma ^{ab}$:
\begin{eqnarray}
S^{x} &=&\frac{\sqrt{3}}{2}\Gamma ^{34}+\frac{1}{2}(\Gamma ^{12}+\Gamma
^{35}),  \notag \\
S^{y} &=&-\frac{\sqrt{3}}{2}\Gamma ^{24}-\frac{1}{2}(\Gamma ^{13}-\Gamma
^{25}),\text{ }  \notag \\
S^{z} &=&-\Gamma ^{15}+\frac{1}{2}\Gamma ^{23},
\end{eqnarray}%
which form the spin SU(2) algebra with $[S^{\alpha },S^{\beta }]=i\epsilon
_{\alpha \beta \gamma }S^{\gamma }$. Moreover, there are seven spin-octupole
matrices given by%
\begin{eqnarray}
(S^{x})^{3} &=&\frac{7\sqrt{3}}{8}\Gamma ^{34}+\frac{7}{8}\Gamma ^{35}+\frac{%
13}{8}\Gamma ^{12},  \notag \\
(S^{y})^{3} &=&-\frac{7\sqrt{3}}{8}\Gamma ^{24}+\frac{7}{8}\Gamma ^{25}-%
\frac{13}{8}\Gamma ^{13},  \notag \\
(S^{z})^{3} &=&\frac{13}{8}\Gamma ^{23}-\frac{7}{4}\Gamma ^{15},  \notag \\
\{S^{x},(S^{y})^{2}-(S^{z})^{2}\} &=&-\frac{\sqrt{3}}{2}\Gamma ^{34}+\frac{3%
}{2}\Gamma ^{35},  \notag \\
\{S^{y},(S^{z})^{2}-(S^{x})^{2}\} &=&-\frac{\sqrt{3}}{2}\Gamma ^{24}-\frac{3%
}{2}\Gamma ^{25},  \notag \\
\{S^{z},(S^{x})^{2}-(S^{y})^{2}\} &=&\sqrt{3}\Gamma ^{14},  \notag \\
S^{x}S^{y}S^{z}+S^{z}S^{y}S^{x} &=&\frac{\sqrt{3}}{2}\Gamma ^{45},
\end{eqnarray}%
where we have used the abbreviation $\left\{ A,B\right\} =AB+BA$. The
spin-quadrupole operators, spin-dipole and spin-octupole operators can be
simply expressed as
\begin{eqnarray}
n_{i}^{a} &=&\frac{1}{2}\sum_{\alpha \beta }\psi _{i\alpha }^{\dag }\Gamma
_{\alpha \beta }^{a}\psi _{i\beta }\;(1\leq a\leq 5), \\
L_{i}^{ab} &=&-\frac{1}{2}\sum_{\alpha \beta }\psi _{i\alpha }^{\dag }\Gamma
_{\alpha \beta }^{ab}\psi _{i\beta }\;(1\leq a<b\leq 5),
\end{eqnarray}%
where $n_{i}^{a}$ corresponds to an SO(5) superspin vector, and $L_{i}^{ab}$
constitute the adjoint representation of the SO(5) Lie group satisfying the
commutation relations:%
\begin{equation}
\lbrack L_{i}^{ab},L_{j}^{cd}]=i\delta _{ij}(\delta _{ad}L_{i}^{bc}+\delta
_{bc}L_{i}^{ad}-\delta _{ac}L_{i}^{bd}-\delta _{bd}L_{i}^{ac}).
\end{equation}%
Furthermore, $n_{i}^{a}$ and $L_{i}^{ab}$ together as generators form an
SU(4) Lie group, the highest symmetry group for four-component spin-$\frac{3%
}{2}$ fermions in the particle-hole channel. Under the SO(5)
transformations, the sixteen bilinear operators in the particle-hole channel
are classified as particle number operator (scalar), spin-quadrupole
operators (vector), spin-dipole and spin-octupole operators (tensor).

We would like to emphasize that in the multipolar expansion of classical
electrodynamics, the higher-order multipolar interactions are usually much
weaker than the dipolar interactions. However, this is not the case in the
present system, where spin dipolar and multipolar operators are all
expressed in terms of bilinear combinations of the fermionic operators.

In this paper, we will focus on the Mott insulating phases with two atoms
per site. In the limit of $t=0$, each lattice site decouples from its
nearest neighbor sites, then the single site ground state is a spin singlet (%
$S_{i}=0$) for $c_{2}>0$ and a spin quintet ($S_{i}=2$) for $c_{2}<0$. For $%
c_{2}=0$, the ground state is a six-fold degenerate state and the
interactions gain an enhanced SU(4) symmetry. Since the scattering lengths
for the spin-$\frac{3}{2}$ fermionic ultracold atoms are not available yet,
we consider the parameter regime of $c_{0}\gg |c_{2}|,t$, where the
spin-dependent interaction are much weaker than the spin-independent one.
Actually, this request is well satisfied for current alkali atom
experiments. Under these considerations, the stability of Mott insulator
with double occupancy is also guaranteed and an interesting phase diagram
with three different kinds of Mott insulating states will be obtained.

\section{Spin-quadrupole ordering state}

In this section, we briefly review the Mott insulating phase with staggered
spin-quadrupole ordering formed by the half-filled spin-$\frac{3}{2}$
fermionic atoms in a two-dimensional optical lattice \cite{Tu-2006}. To
reveal such an order, the generalized Hubbard model can be rewritten in an
SO(5) invariant form \cite{CWu-2003}%
\begin{eqnarray}
H &=&-t\sum_{\langle ij\rangle ,\alpha }(\psi _{i\alpha }^{\dag }\psi
_{j\alpha }+\text{H.c.})-\frac{3c_{2}}{4}\sum_{i,1\leq a\leq
5}(n_{i}^{a})^{2}  \notag \\
&&+\frac{8c_{0}-15c_{2}}{16}\sum_{i}(N_{i}-2)^{2}-\mu \sum_{i}N_{i}.
\end{eqnarray}%
Provided the commutation rules%
\begin{equation}
\lbrack L_{i}^{ab},n_{j}^{c}]=i\delta _{ij}(\delta _{bc}n_{i}^{a}-\delta
_{ac}n_{i}^{b})
\end{equation}%
are satisfied, the generic SO(5) symmetry of this model manifests
explicitly, i.e., the Hamiltonian is invariant under the transformation%
\begin{equation}
\psi \rightarrow \psi ^{\prime }=e^{-iL^{ab}\theta _{ab}}\psi ,\text{ }1\leq
a<b\leq 5,
\end{equation}%
where $\theta _{ab}$ are the Euler angles representing infinitesimal
rotations between the ten independent planes in five dimensions. At half
filling, the chemical potential $\mu $ is zero and the average number of
fermions per site is $\langle N_{i}\rangle =2$. As far as an insulating
state is concerned, the particle number fluctuations can be neglected and we
focus on the quantum spin fluctuations, which result in various spin
ordering states. In particular, for $c_{2}>0$, site-singlets formed by two
spin-$\frac{3}{2}$ fermions with total spin $S=0$ are energetically
favorable. In this case, the quantum spin-quadrupole fluctuations are an
important feature of the generalized Hubbard model. As we have shown in the
previous work \cite{Tu-2006}, the doubly occupied fermionic atoms in a
square lattice can exhibit a staggered spin-quadrupole ordering with the
order parameter $(-1)^{i}\langle n_{i}^{a}\rangle \neq 0$. Without loss of
generality, the order parameter can be chosen as $(-1)^{i}\langle
n_{i}^{4}\rangle \neq 0$, where
\begin{equation}
2n_{i}^{4}=\psi _{i,\frac{3}{2}}^{\dagger }\psi _{i,\frac{3}{2}}+\psi _{i,-%
\frac{3}{2}}^{\dagger }\psi _{i,-\frac{3}{2}}-\psi _{i,\frac{1}{2}}^{\dagger
}\psi _{i,\frac{1}{2}}-\psi _{i,-\frac{1}{2}}^{\dagger }\psi _{i,-\frac{1}{2}%
},
\end{equation}%
corresponding to the difference between the $S^{z}=\pm 3/2$ and $S^{z}=\pm
1/2$ spin densities in the site singlet state. The typical configuration of
this particular spin-quadrupole ordering state is shown in Fig. 1. This
quantum analogue of liquid crystal state breaks the spin SU(2) rotational
symmetry, but preserves the time reversal symmetry.

Due to the fixed direction $n_{i}^{4}$ of the SO(5) superspin vector, the
symmetry of the spin-quadrupole ordering ground state is spontaneously
broken down to SO(4) and the Goldstone manifold is SO(5)/SO(4)=S$^{4}$,
which is a four-dimensional sphere. Since $L_{i}^{ab}$ performs a rotation
between $n_{i}^{a}$ and $n_{i}^{b}$, the generators of the SO(4) symmetry
are $L_{i}^{12}$, $L_{i}^{13}$, $L_{i}^{15}$, $L_{i}^{23}$, $L_{i}^{25}$,
and $L_{i}^{35}$. Thus, there are four Goldstone bosonic modes without
coupling to each other due to the residual SO(4) symmetry. The presence of
this high symmetry qualitatively affects the strong coupling behavior of the
spin-quadrupole density waves, whose velocity is saturated at large $c_{2}$
as $v_{Q}\sim t$. However, in the antiferromagnetic ordering state of the
half-filled spin-$\frac{1}{2}$ Hubbard model, the Goldstone manifold is
SO(3)/SO(2)=S$^{2}$. The residual SO(2) symmetry does not forbid transverse
spin wave mixing, and the density wave velocity is suppressed as $v_{\text{%
SDW}}\sim t^{2}/U$ in the strong coupling limit.

\begin{figure}[tbp]
\centering \includegraphics [width=4.5cm]{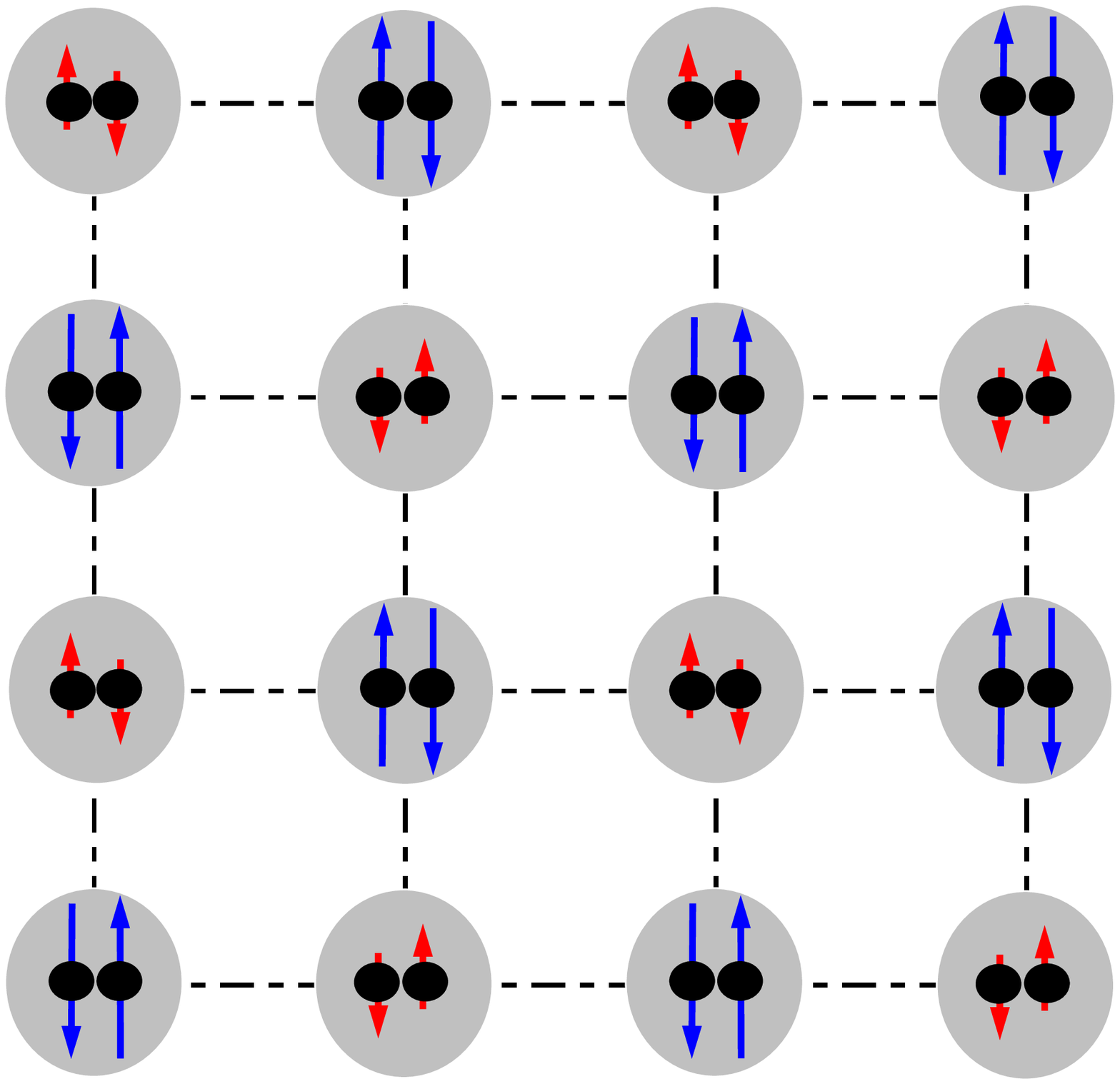}
\caption{(Color online) The picture of the staggered spin-quadrupole
long-range ordered state with the order parameter $(-1)^{i}\langle
n_{i}^{4}\rangle $ on a two-dimensional square lattice, where each site is
occupied by a spin singlet state. The long blue arrow and short red arrow
denote a fermion with $S^{z}=\pm \frac{3}{2}$ and $S^{z}=\pm \frac{1}{2}$,
respectively.}
\end{figure}

Another interesting physics in the spin-quadrupole ordering state is the
non-abelian Berry phase. Since the four-component SO(5) spinor forms a
seven-dimensional sphere $S^{7}$, the adiabatic rotation of the fermionic
quasiparticle in spin-quadrupole ordering state defines\ the second Hopf map%
\begin{equation}
S^{7}\rightarrow S^{4},
\end{equation}%
which enables an interesting non-abelian SU(2) gauge connection of the
so-called Wilczek-Zee holonomy \cite{Wilczek-1984,Demler-1999}.
Mathematically, it is understood that $S^{7}$ is an SU(2) bundle over $S^{4}$
sphere. Therefore, the topological structure of the spin-quadrupole ordering
state is described by the second Hopf map and the second Chern number. As
the fermionic quasiparticle spectrum is two-fold degenerate, once we perform
an adiabatic circuit rotation of the superspin vector $n_{i}^{a}$, the
spinor state can be a linear combination of the two degenerate low energy
states, which is related to the original state by a $2\times 2$ unitary
transformation matrix. Actually, the SU(2) non-abelian Berry phase is also
present in the quintet pairing state \cite{Chern-2004} and plays a key role
in the entanglement generation between the fermionic quasiparticles and the
half-quantum vortex \cite{CWu-Hu-2005}.

\section{Degenerate antiferromagnetic spin-dipole / spin-octupole ordering
state}

\subsection{Saddle point approximation}

Next we will consider the case for $c_{2}<0$, where a spin quintet state is
formed by two spin-$\frac{3}{2}$ fermions on each lattice site with total
spin $S=2$. Applying the SU(4) Fierz identity%
\begin{equation}
\sum_{1\leq a<b\leq 5}(L_{i}^{ab})^{2}+\sum_{1\leq a\leq 5}(n_{i}^{a})^{2}+%
\frac{5}{4}(N_{i}-2)^{2}=5,
\end{equation}%
to the interactions of the model Hamiltonian, we can rewrite the generalized
one-band Hubbard model as%
\begin{eqnarray}
H &=&-t\sum_{\langle ij\rangle ,\alpha }(\psi _{i\alpha }^{\dag }\psi
_{j\alpha }+\text{H.c.})+\frac{3c_{2}}{4}\sum_{i,1\leq a<b\leq
5}(L_{i}^{ab})^{2}  \notag \\
&&+\frac{c_{0}}{2}\sum_{i}(N_{i}-2)^{2}-\mu \sum_{i}N_{i},
\end{eqnarray}%
where the chemical potential $\mu $ is set zero at half filling. As far as
the Mott insulating phase is concerned, the particle number fluctuations can
be neglected safely, and the spin-dipole and spin-octupole fluctuations
become dominant when $c_{2}<0$.

Formally, the partition function is written as a path integral
\begin{equation}
Z=\int \mathcal{D}\psi ^{\dag }\mathcal{D}\psi \exp \left[ -\int_{0}^{\beta
}L\left( \tau \right) d\tau \right] ,
\end{equation}%
where $\beta =\frac{1}{k_{B}T}$ and the Lagrangian is given by $%
L=\sum_{i,\alpha }\psi _{i\alpha }^{\dag }\partial _{\tau }\psi _{i\alpha
}+H $. In the following we denote $U\equiv -\frac{3}{4}c_{2}$ for simplicity
and a Hubbard-Stratonovich transformation is performed%
\begin{eqnarray*}
&&\exp \left[ U\int_{0}^{\beta }d\tau \sum_{i,a<b}(L_{i}^{ab})^{2}\right] \\
&=&\int \mathcal{D}\varphi \exp \left[ \int_{0}^{\beta }d\tau
\sum_{i,a<b}\left( \sqrt{2U}\varphi _{i}^{ab}L_{i}^{ab}-\frac{1}{2}(\varphi
_{i}^{ab})^{2}\right) \right] ,
\end{eqnarray*}%
where a ten-component order parameter field $\varphi $ is introduced. By
integrating out the fermion fields, an effective action is obtained as%
\begin{equation}
S_{\text{eff}}=\int_{0}^{\beta }d\tau \sum_{i,a<b}\frac{1}{2}(\varphi
_{i}^{ab})^{2}-\text{Tr}\ln \left[ \partial _{\tau }+\mathbf{M}\right] ,
\end{equation}%
where the trace is taken over the spinor space, the spatial and imaginary
time coordinates. The matrix element of $\mathbf{M}$ and Green's function
(GF) of the fermions satisfy
\begin{eqnarray}
\langle \mathbf{r}_{i},\tau ,\alpha |\mathbf{M}|\mathbf{r}_{j},\tau ^{\prime
},\beta \rangle &=&-2t\delta _{\tau \tau ^{\prime }}\delta _{i,j+\delta
}\delta _{\alpha \beta }  \notag \\
&&+\sqrt{\frac{U}{2}}\delta _{\tau \tau ^{\prime }}\delta
_{ij}\sum_{a<b}\varphi _{i}^{ab}\Gamma _{\alpha \beta }^{ab},  \notag \\
G_{\alpha \beta }(\mathbf{r}_{i},\tau ;\mathbf{r}_{j},\tau ^{\prime })
&=&\langle \mathbf{r}_{i},\tau ,\alpha |\frac{-1}{\partial _{\tau }+\mathbf{M%
}}|\mathbf{r}_{j},\tau ^{\prime },\beta \rangle .
\end{eqnarray}%
So far no approximations have been made. Once we perform a loop expansion on
the effective action with respect to $\varphi $, the instability owing the
negative coefficient of the quadratic term in $\varphi $ is dominated by the
so-called Fermi surface instability, which is similar to the well-known
antiferromagnetic instability in half filled Hubbard model \cite{Nagaosa}.

In order to determine the long range order of Mott insulator, we first
consider the saddle point solution of the effective action. Differentiating $%
S_{\text{eff}}$ with respect to $\varphi _{i}^{ab}\left( \tau \right) $, we
obtain the equation%
\begin{equation}
\varphi _{i}^{ab}\left( \tau \right) =-\sqrt{\frac{U}{2}}\underset{\alpha
\beta }{\sum }G_{\alpha \beta }(\mathbf{r}_{i},\tau ;\mathbf{r}_{i},\tau
)\Gamma _{\beta \alpha }^{ab}.
\end{equation}%
Then it is expected that the ground state is in a staggered phase of the
SO(5) adjoint order parameter, namely, a staggered degenerate spin-dipole
/spin-octupole ordering phase with a static order parameter field $\varphi
_{i}^{ab}\left( \tau \right) \rightarrow |\varphi |e^{i\mathbf{Q}\cdot
\mathbf{r}_{i}}d^{ab}$, where $\mathbf{Q}=(\pi ,\pi )$ corresponds to the
nesting vector in a two-dimensional square lattice, and $d^{ab}$ is the
symmetry breaking direction. The degeneracy of spin-dipole and spin-octupole
order is protected by the exact SO(5) symmetry of the Hamiltonian, which
implies the spin-dipole operators can be rotated to spin-octupole operators
and vice versa. Thus, this degeneracy is robust against quantum and thermal
fluctuations. However, an anisotropy, such as lattice potential, will pin
down a special choice of the ground state. Actually, the effective action at
the saddle point becomes
\begin{eqnarray}
S^{(0)} &=&\sum_{\mathbf{k},i\mathbf{\omega }_{n}}\left\{ \psi _{\mathbf{k}%
}^{\dag }(i\mathbf{\omega }_{n})\left( -i\mathbf{\omega }_{n}+\varepsilon _{%
\mathbf{k}}\right) \psi _{\mathbf{k}}(i\mathbf{\omega }_{n})+\frac{1}{2}%
\beta N|\varphi |^{2}\right.  \notag \\
&&\left. +\sqrt{\frac{U}{2}}|\varphi |\sum_{a<b}\psi _{\mathbf{k}}^{\dag }(i%
\mathbf{\omega }_{n})d^{ab}\Gamma ^{ab}\psi _{\mathbf{k}-\mathbf{Q}}(i%
\mathbf{\omega }_{n})\right\} ,
\end{eqnarray}%
where the fermion dispersion relation is $\varepsilon _{\mathbf{k}}=-2t(\cos
k_{x}+$ $\cos k_{y})$, and $\mathbf{\omega }_{n}$ is the fermionic Matsubara
frequency. The fermion GF has to be defined as $G_{\alpha \beta }\left(
\mathbf{k},\mathbf{k}^{\prime };\tau \right) =-\langle T_{\tau }\psi _{%
\mathbf{k}\alpha }(\tau )\psi _{-\mathbf{k}^{\prime }\beta }^{\dag
}(0)\rangle $, which has non-zero off-diagonal terms in momentum space due
to the umklapp processes with respect to $\mathbf{Q}$. At the saddle point,
the fermion single particle GF is given by%
\begin{eqnarray}
&&G_{\alpha \beta }\left( \mathbf{k},-\mathbf{k}^{\prime };i\omega
_{n}\right)  \notag \\
&=&\frac{\left( i\omega _{n}+\varepsilon _{\mathbf{k}}\right) \delta
_{\alpha \beta }\delta _{\mathbf{kk}^{\prime }}+\sqrt{\frac{U}{2}}|\varphi
|\sum_{a<b}d^{ab}\Gamma _{\alpha \beta }^{ab}\delta _{\mathbf{k}^{\prime },%
\mathbf{k}-\mathbf{Q}}}{(i\omega _{n})^{2}-E_{\mathbf{k}}^{2}},
\end{eqnarray}%
where GF poles determine the fermionic quasiparticle spectrum $E_{\mathbf{k}%
}=\pm \sqrt{\varepsilon _{\mathbf{k}}^{2}+\frac{U}{2}|\varphi |^{2}}$. At
half filling, the upper band is empty while the lower band\ is completely
filled. Thus an energy gap $\Delta =2\sqrt{\frac{U}{2}}|\varphi |$ opens up
in the single particle spectrum. At $T=0$K, the gap equation is given by%
\begin{equation}
\frac{1}{U}=\frac{1}{N}\sum_{\mathbf{k}}\frac{1}{E_{\mathbf{k}}}.
\end{equation}%
In the limit of $U\ll t$, it gives rise to $\Delta \simeq 2te^{-\pi \sqrt{%
2t/U}}$, which implies the Fermi surface is fully gapped for arbitrary small
but finite $U$. For $U\gg t$, we have $\Delta \simeq 2U$, i.e., a large Mott
gap.

We have shown two typical configurations of this long-range ordering phase
in Fig. 2, and the corresponding order parameters are denoted by $%
(-1)^{i}\langle L_{i}^{15}\rangle \neq 0$ and $(-1)^{i}\langle
L_{i}^{23}\rangle \neq 0$, respectively. More explicitly, $L_{i}^{15}$ and $%
L_{i}^{23}$ are expressed as
\begin{eqnarray*}
2L_{i}^{15} &=&\psi _{i,\frac{3}{2}}^{\dagger }\psi _{i,\frac{3}{2}}+\psi
_{i,\frac{1}{2}}^{\dagger }\psi _{i,\frac{1}{2}}-\psi _{i,-\frac{1}{2}%
}^{\dagger }\psi _{i,-\frac{1}{2}}-\psi _{i,-\frac{3}{2}}^{\dagger }\psi
_{i,-\frac{3}{2}}, \\
2L_{i}^{23} &=&\psi _{i,\frac{1}{2}}^{\dagger }\psi _{i,\frac{1}{2}}+\psi
_{i,-\frac{3}{2}}^{\dagger }\psi _{i,-\frac{3}{2}}-\psi _{i,\frac{3}{2}%
}^{\dagger }\psi _{i,\frac{3}{2}}-\psi _{i,-\frac{1}{2}}^{\dagger }\psi _{i,-%
\frac{1}{2}},
\end{eqnarray*}%
corresponding to two different patterns of total $S^{z}=\pm 2$ and $%
S^{z}=\pm 1$ of two atoms in the site-quintet state, respectively. Actually,
both configurations display the coexistence of antiferromagnet spin-dipole
and staggered spin-octupole ordering with breaking the spin SU(2) and time
reversal symmetries.

\begin{figure}[tbp]
\hspace{0.1cm} \includegraphics [width=3.5cm]{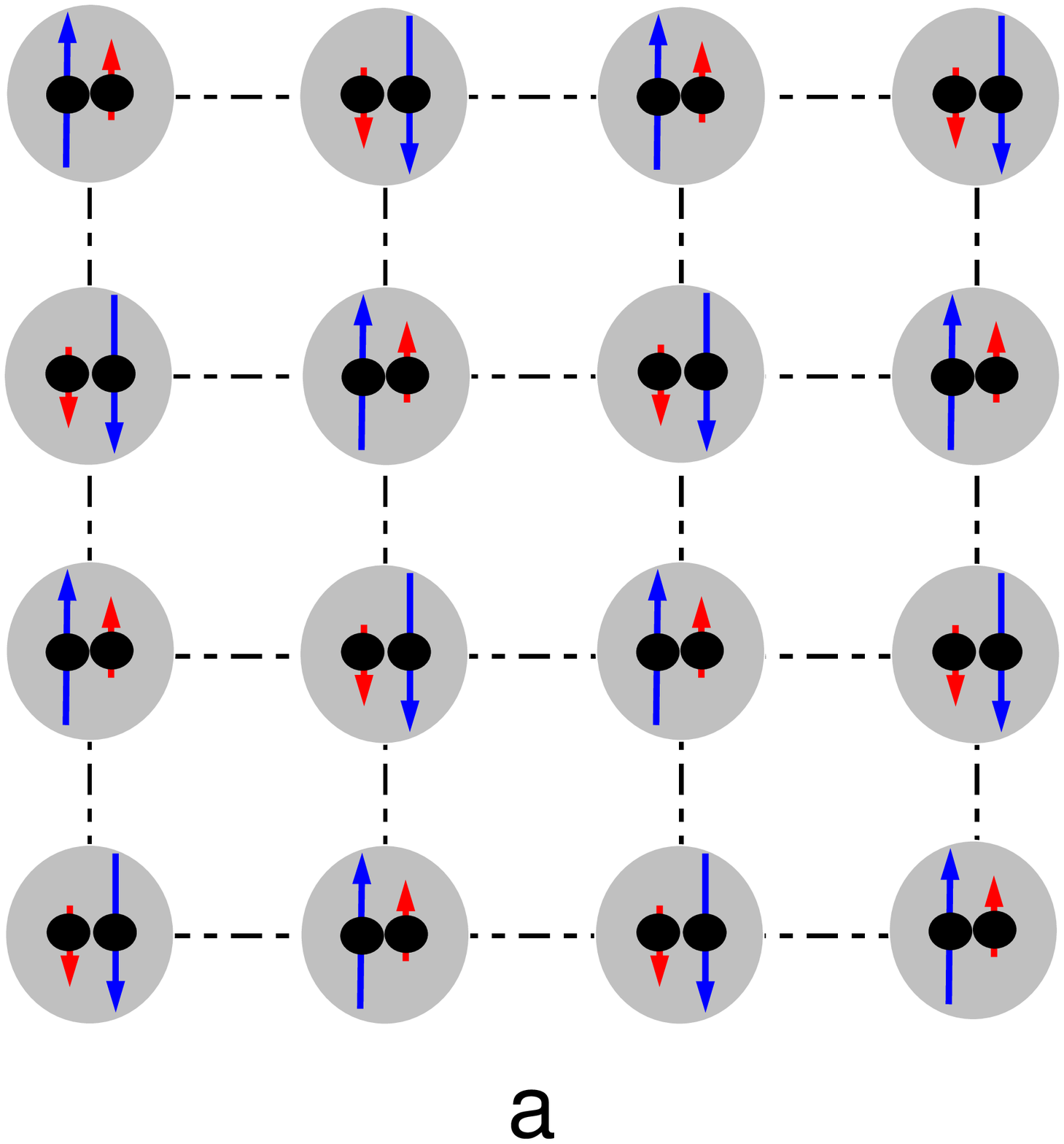} \hspace{0.5cm}
\includegraphics
[width=3.5cm]{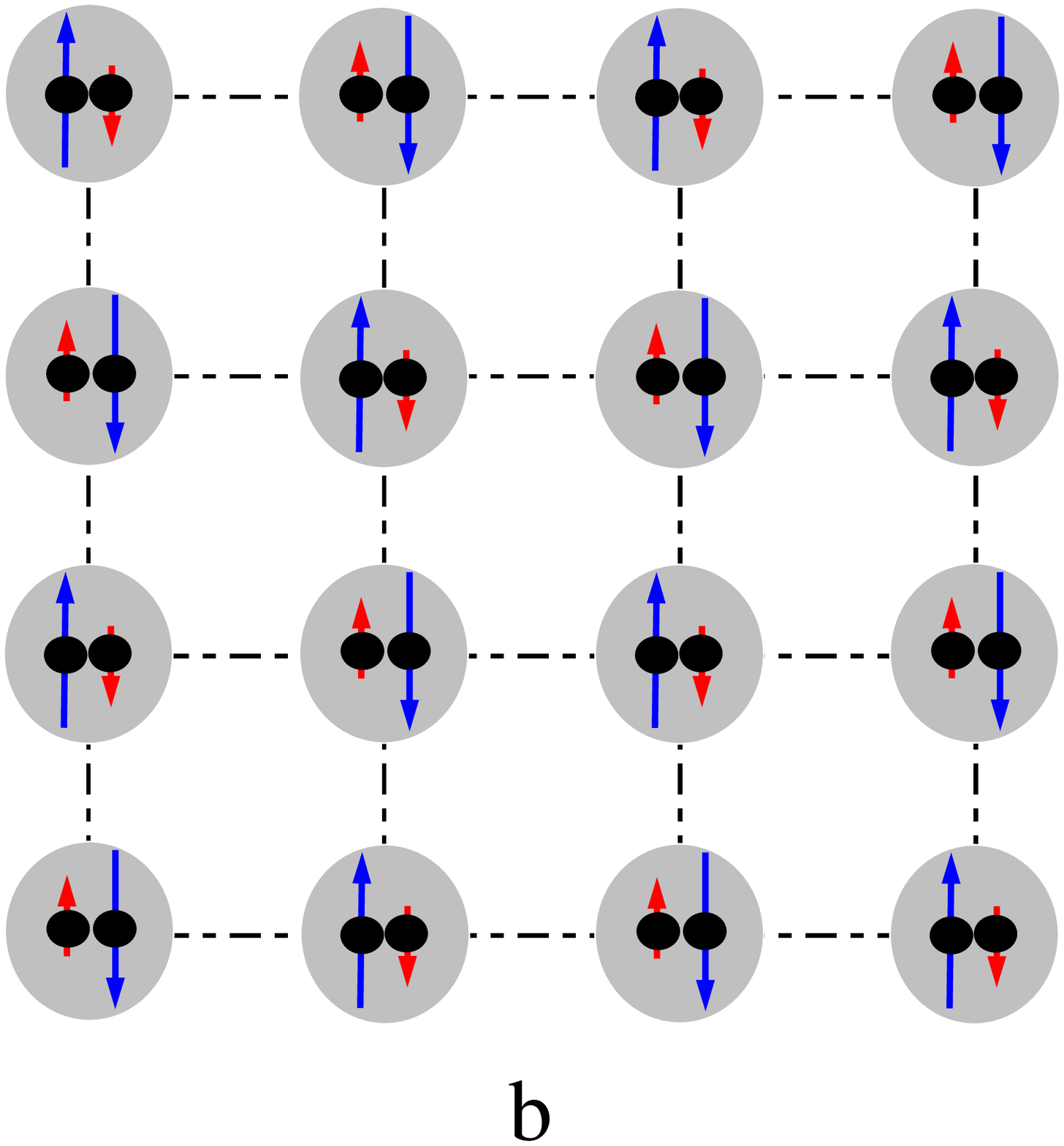}
\caption{(Color online) The picture of the long range ordered state with the
order parameter $(-1)^{i}\langle L_{i}^{15}\rangle $ (a) and $%
(-1)^{i}\langle L_{i}^{23}\rangle $ (b) on a two-dimensional square lattice,
where each site is occupied by a spin quintet state.}
\end{figure}

To simplify the following discussions, we assume the order parameter is
fixed by $(-1)^{i}\langle L_{i}^{15}\rangle \neq 0$. Along with the order
parameter $L_{i}^{15}$ acquires non-zero vacuum expectation value, the SO(5)
generators $L_{i}^{15}$, $L_{i}^{23}$, $L_{i}^{24}$ and $L_{i}^{34}$ still
keep the ordering state invariant, where the latter three SO(5) generators
form an SU(2) subgroup. Therefore, the Goldstone manifold is SO(5)/[SU(2)$%
\times $U(1)]=CP$^{3}$, which is a six-dimensional complex projective space.
According to the Goldstone theorem, there will be six branches of Goldstone
bosons induced by the spontaneous symmetry breaking.

The topological aspects on the degenerate spin-dipole/spin-octupole ordering
phase are also interesting. Mathematically, a seven-dimensional sphere $%
S^{7} $ can be viewed as a U(1) bundle over CP$^{3}$, which is the Goldstone
manifold in the current case. Because the four-component SO(5) spinor lives
on $S^{7}$, the adiabatic circuit rotation of the fermionic quasiparticles
in the degenerate spin-dipole with spin-octupole ordering state defines a
U(1) gauge connection. Although the fermionic quasiparticle spectrum is also
two-fold degenerate, once an adiabatic circuit rotation of the order
parameter $L_{i}^{ab}$ is performed, only an additional U(1) phase factor
can be generated in the resulting spinor state, instead of linear
combination of the two degenerate states due to the abelian nature of the
U(1) Berry phase \cite{Berry-1984}. Therefore, unlike the spin-quadrupole
ordering state, the topological structure of the degenerate spin-dipole
/spin-octupole ordering state is characterized by the U(1) abelian Berry
phase \ and the first Chern number.

\subsection{Gaussian fluctuations}

To further study the collective excitations, we consider Gaussian
fluctuations around the saddle point solution, $\varphi =\varphi _{c}+\delta
\varphi $. Then the effective action $S_{\text{eff}}$ can be expanded as $S_{%
\text{eff}}=\sum_{n=0}^{\infty }S^{(n)}(\varphi _{c},\delta \varphi )$ from%
\begin{equation}
\text{Tr}\ln \left[ \partial _{\tau }+\mathbf{M}\right] =\text{Tr}\ln (-%
\mathbf{G}^{-1})-\sum_{n}\frac{1}{n}\text{Tr}(\mathbf{GV})^{n},
\end{equation}%
where $\mathbf{G}$ represents the fermion GF at the saddle point and the
matrix element of $\mathbf{V}$ is given by%
\begin{equation}
\langle \mathbf{r}_{i},\tau ,\alpha |\mathbf{V}|\mathbf{r}_{j},\tau ^{\prime
},\beta \rangle =\delta _{\tau \tau ^{\prime }}\delta _{ij}\sqrt{\frac{U}{2}}%
\sum_{a<b}\delta \varphi _{i}^{ab}\left( \tau \right) \Gamma _{\alpha \beta
}^{ab}.
\end{equation}%
Since $\mathbf{V}$ only contains a linear term in $\delta \varphi $, the
above procedure is indeed an expansion in the spin-dipole and spin-octupole
fluctuations. The first order term in $\delta \varphi $ vanishes due to the
saddle point equation. The Fourier transformation of $\delta \varphi
_{i}^{ab}\left( \tau \right) $ is defined by%
\begin{equation}
\delta \varphi _{i}^{ab}\left( \tau \right) =\frac{1}{\sqrt{\beta N}}\sum_{%
\mathbf{p},i\omega _{l}}\delta \varphi _{\mathbf{p}}^{ab}\left( i\omega
_{l}\right) e^{i\mathbf{p}\cdot \mathbf{r}_{i}-i\omega _{l}\tau },
\end{equation}%
where $\omega _{l}$ are bosonic Matsubara frequencies. After some
straightforward algebra, up to the second order expansion in the
fluctuation field $\delta \varphi $, we arrive at
\begin{widetext}
\begin{eqnarray}
S^{(2)} &=&\sum_{\mathbf{p},i\omega _{l}}\left\{ K_{0}(\mathbf{p},i\omega
_{l})\Phi _{-\mathbf{p}}^{T}(-i\omega _{l})\Phi _{\mathbf{p}}(i\omega
_{l})\right.   \notag \\
&&+\sum_{a=2,3,4}K_{1}(\mathbf{p},i\omega _{l})[\delta \varphi _{-\mathbf{p}%
}^{1a}\left( -i\omega _{l}\right) \delta \varphi _{\mathbf{p}}^{1a}\left(
i\omega _{l}\right) +\delta \varphi _{-\mathbf{p}}^{a5}\left( -i\omega
_{l}\right) \delta \varphi _{\mathbf{p}}^{a5}\left( i\omega _{l}\right) ]
\notag \\
&&+\sum_{a=2,3,4}\left. K_{2}(\mathbf{p},i\omega _{l})[\delta \varphi _{-%
\mathbf{p}-\mathbf{Q}}^{1a}\left( -i\omega _{l}\right) \delta \varphi _{%
\mathbf{p}}^{a5}\left( i\omega _{l}\right) -\delta \varphi _{-\mathbf{p}-%
\mathbf{Q}}^{a5}\left( -i\omega _{l}\right) \delta \varphi _{\mathbf{p}%
}^{1a}\left( i\omega _{l}\right) ]\right\} ,  \label{GF2}
\end{eqnarray}%
where $\Phi _{i}=(\delta \varphi _{i}^{15},\delta \varphi _{i}^{23},\delta
\varphi _{i}^{24},\delta \varphi _{i}^{34})^{T}$ is a Nambu spinor and the
kernel functions are given by
\begin{eqnarray}
&&K_{0}(\mathbf{p},i\omega _{l})=\frac{1}{2}-\frac{U}{\beta N}\sum_{\mathbf{k%
},i\mathbf{\omega }_{n}}\frac{\omega _{n}\left( \omega _{n}+\omega
_{l}\right) -\varepsilon _{\mathbf{k}}\varepsilon _{\mathbf{k+p}}-\frac{%
\Delta ^{2}}{4}}{(\omega _{n}^{2}+E_{\mathbf{k}}^{2})\left[ \left( \omega
_{n}+\omega _{l}\right) ^{2}+E_{\mathbf{k}+\mathbf{p}}^{2}\right] }, \\
&&K_{1}(\mathbf{p},i\omega _{l})=\frac{1}{2}-\frac{U}{\beta N}\sum_{\mathbf{k%
},i\mathbf{\omega }_{n}}\frac{\omega _{n}\left( \omega _{n}+\omega
_{l}\right) -\varepsilon _{\mathbf{k}}\varepsilon _{\mathbf{k+p}}+\frac{%
\Delta ^{2}}{4}}{(\omega _{n}^{2}+E_{\mathbf{k}}^{2})\left[ \left( \omega
_{n}+\omega _{l}\right) ^{2}+E_{\mathbf{k}+\mathbf{p}}^{2}\right] }, \\
&&K_{2}(\mathbf{p},i\omega _{l})=\frac{U\Delta }{2\beta N}\sum_{\mathbf{k},i%
\mathbf{\omega }_{n}}\frac{\omega _{l}}{(\omega _{n}^{2}+E_{\mathbf{k}}^{2})%
\left[ \left( \omega _{n}+\omega _{l}\right) ^{2}+E_{\mathbf{k}+\mathbf{p}%
}^{2}\right] }.
\end{eqnarray}%
In the low energy limit $i\omega _{l}\rightarrow 0$, for the momentum
transfer $\mathbf{p}=\mathbf{Q}$, $K_{0}\left( \mathbf{Q},0\right) >0$ and $%
K_{1}\left( \mathbf{Q},0\right) =K_{2}\left( \mathbf{Q},0\right) =0$
following from the gap equation. Therefore, the collective excitation modes
in the first term of Eq.(\ref{GF2}) is gapped, and the remaining terms
describe the Goldstone modes related to the symmetry breaking. From Eq.(\ref%
{GF2}), we notice that the couplings between the fluctuating Goldstone
fields are in a twin form as $\delta \varphi ^{12}\delta \varphi ^{25}$, $%
\delta \varphi ^{13}\delta \varphi ^{35}$ and $\delta \varphi ^{14}\delta
\varphi ^{45}$. As we discussed, the resulting effective action enjoys a
residual SU(2)$\times $U(1) symmetry and the coupling between the Goldstone
fields are allowed.

In order to evaluate the low energy behavior of the Goldstone bosons, we
perform the summation over the Matsubara frequency in the kernel functions
of $K_{1}$, $K_{2}$ and analytical continuation $i\omega _{l}\rightarrow
\omega +i0^{+}$. When the long wavelength and low energy limit are
considered at $T=0$K, the kernel functions are expanded to the leading order
in $\mathbf{q}$ and $\omega $ as
\begin{eqnarray}
K_{1}(\mathbf{Q}+\mathbf{q},\omega ) &\approx &a\mathbf{q}^{2}-b\omega ^{2},%
\text{ }K_{1}(\mathbf{q},\omega )\approx d,  \notag \\
K_{2}(\mathbf{Q}+\mathbf{q},\omega ) &=&K_{2}(\mathbf{q},\omega )\approx
c\omega ,
\end{eqnarray}%
where the coefficients are given by%
\begin{equation}
a=\int_{-\pi }^{\pi }\frac{d^{2}\mathbf{k}}{(2\pi )^{2}}\frac{Ut^{2}\sin
^{2}k_{x}}{2E_{\mathbf{k}}^{3}},\text{ } b=\int_{-\pi }^{\pi }\frac{d^{2}%
\mathbf{k}}{(2\pi )^{2}}\frac{U}{8E_{\mathbf{k}}^{3}}, c=-\int_{-\pi }^{\pi }%
\frac{d^{2}\mathbf{k}}{(2\pi )^{2}}\frac{iU\Delta }{8E_{\mathbf{k}}^{3}},%
\text{ } d=\int_{-\pi }^{\pi }\frac{d^{2}\mathbf{k}}{(2\pi )^{2}}\frac{%
U\varepsilon _{\mathbf{k}}^{2}}{2E_{\mathbf{k}}^{3}}.
\end{equation}
Thus, the effective action describing the Goldstone bosons is given by%
\begin{equation}
S_{\text{GB}}=\int \frac{d\mathbf{q}d\omega }{(2\pi )^{3}}%
\sum_{a=2,3,4}\left(
\begin{array}{c}
\delta \varphi _{-\mathbf{q}-\mathbf{Q}}^{1a}\left( -\omega \right) \\
\delta \varphi _{-\mathbf{q}}^{1a}\left( -\omega \right) \\
\delta \varphi _{-\mathbf{q}-\mathbf{Q}}^{a5}\left( -\omega \right) \\
\delta \varphi _{-\mathbf{q}}^{a5}\left( -\omega \right)%
\end{array}%
\right) ^{T} \left(
\begin{array}{cccc}
a\mathbf{q}^{2}-b\omega ^{2} & 0 & 0 & c\omega \\
0 & d & c\omega & 0 \\
0 & -c\omega & a\mathbf{q}^{2}-b\omega ^{2} & 0 \\
-c\omega & 0 & 0 & d%
\end{array}%
\right) \left(
\begin{array}{c}
\delta \varphi _{\mathbf{q}+\mathbf{Q}}^{1a}\left( \omega \right) \\
\delta \varphi _{\mathbf{q}}^{1a}\left( \omega \right) \\
\delta \varphi _{\mathbf{q}+\mathbf{Q}}^{a5}\left( \omega \right) \\
\delta \varphi _{\mathbf{q}}^{a5}\left( \omega \right)%
\end{array}%
\right) .
\end{equation}%
\end{widetext}
Then the dispersion relation of the Goldstone bosons is determined by the
pole of the collective mode correlation functions, giving rise to $\omega =v|%
\mathbf{q}|$, a linear dispersion with the density wave velocity%
\begin{equation}
v=\sqrt{\frac{ad}{bd-c^{2}}}.
\end{equation}%
The corresponding numerical results are displayed in Fig. 3.

\begin{figure}[tbp]
\includegraphics[scale=0.8]{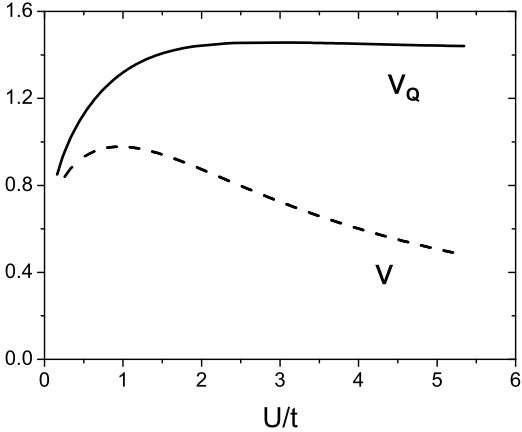}
\caption{The degenerate spin-dipole/spin-octupole density wave velocity $v$
(dotted line) is compared with the spin-quadrupole wave velocity $v_{\text{Q}%
}$ (solid line).}
\end{figure}

In the small $U$ limit, the density wave velocity is approximated by $%
v\approx 2t(2U/t)^{1/4}/\sqrt{\pi }$, while in the large $U$ limit, it is
given by $v\approx 2\sqrt{2}t^{2}/U$. Due to the presence of the coupled
vibrations of the transverse modes, the wave velocity is suppressed in the
strong coupling limit, in a sharp contrast to the velocity of the
spin-quadrupole density waves. These properties can be understood on the
basis of the symmetry considerations. The transverse mode couplings in
spin-quadrupole density waves are absent because the SO(4) invariance of the
effective action under the Gaussian fluctuations does not allow such
couplings. However, the residual SU(2)$\times $U(1) symmetry does allow
coupled vibrations of the Goldstone modes, which is analogous to the
transverse spin wave mixing in half filled spin-$\frac{1}{2}$ Hubbard model.

\section{Algebraic spin liquid state}

In the previous sections, we have established the spin-ordered Mott
insulating phases of spin-$\frac{3}{2}$ fermions in various limits. Near the
special SU(4) point, the energies of the ordering states are very close to
each other. Since these two competing ordering ground states exhibit quite
different symmetry properties, low energy excitations and topological
characterizations, a continuous quantum phase transition from one to another
is almost \textit{impossible}. Therefore a nontrivial quantum critical state
may emerge around the SU(4) symmetric point.

\subsection{SU(4) symmetric point}

Let us first consider the SU(4) point ($c_{2}=0$). In this case, the
generalized one-band Hubbard model at half filling is reduced to the SU(4)
Hubbard model:%
\begin{equation}
H=-t\sum_{\langle ij\rangle ,\alpha }(\psi _{i\alpha }^{\dag }\psi _{j\alpha
}+\text{H.c.})+\frac{c_{0}}{2}\sum_{i}(N_{i}-2)^{2}.
\end{equation}%
When the strong coupling limit ($c_{0}\gg t$) is considered, the effective
exchange model can be derived by a second-order perturbation theory, which
reads%
\begin{equation}
H_{\text{eff}}=J\sum_{\langle ij\rangle }(\sum_{1\leq a<b\leq
5}L_{i}^{ab}L_{j}^{ab}+\sum_{1\leq a\leq 5}n_{i}^{a}n_{j}^{a}),
\end{equation}%
where the coupling constant $J\sim t^{2}/c_{0}$. When we introduce a spinon
representation in terms of four component fermions, the operators $n_{i}^{a}$
and $L_{i}^{ab}$ can be expressed as%
\begin{eqnarray}
n_{i}^{a} &=&\frac{1}{2}\sum_{\alpha \beta }f_{i\alpha }^{\dag }\Gamma
_{\alpha \beta }^{a}f_{i\beta }\;(1\leq a\leq 5),  \notag \\
L_{i}^{ab} &=&-\frac{1}{2}\sum_{\alpha \beta }f_{i\alpha }^{\dag }\Gamma
_{\alpha \beta }^{ab}f_{i\beta }\;(1\leq a<b\leq 5),
\end{eqnarray}%
with a local constraint $\sum_{\alpha }f_{i\alpha }^{\dag }f_{i\alpha }=2$.
The effective model Hamiltonian can thus be written as%
\begin{equation}
H_{\text{eff}}=-J\sum_{\langle ij\rangle ,\alpha \beta }f_{i\alpha }^{\dag
}f_{j\alpha }f_{j\beta }^{\dag }f_{i\beta }.
\end{equation}%
This is nothing but the SU(4) quantum antiferromagnetic Heisenberg model,
which was studied in the large-$N$ limit on a square lattice by Affleck and
Marston \cite{Affleck-1988}. Formally, we can express the partition function
as a path integral%
\begin{equation}
Z=\int \mathcal{D}f^{\dag }\mathcal{D}f\mathcal{D}\lambda \exp \left[
-\int_{0}^{\beta }L_{\text{eff}}(\tau )d\tau \right] ,
\end{equation}%
where the effective Lagrangian is given by%
\begin{equation}
L_{\text{eff}}=\sum_{i,\alpha }f_{i\alpha }^{\dag }\partial _{\tau
}f_{i\alpha }+H_{\text{eff}}+\sum_{i}\lambda _{i}\left( \sum_{\alpha
}f_{i\alpha }^{\dag }f_{i\alpha }-2\right) ,
\end{equation}%
and the Lagrange multiplier field $\lambda _{i}$ is introduced to impose the
local constraint. Then the interaction part can be decoupled via a valence
bond auxiliary field%
\begin{eqnarray}
&&\exp \left[ \int_{0}^{\beta }d\tau \sum_{\langle ij\rangle ,\alpha \beta
}f_{i\alpha }^{\dag }f_{j\alpha }f_{j\beta }^{\dag }f_{i\beta }\right]
\notag \\
&=&\int \mathcal{D}\chi e^{\int_{0}^{\beta }d\tau \left[ \sum_{\langle
ij\rangle ,\alpha }(f_{i\alpha }^{\dag }f_{j\alpha }\chi _{ji}+\text{c.c.}%
)-\sum_{\langle ij\rangle }|\chi _{ij}|^{2}\right] },
\end{eqnarray}%
where $\chi _{ij}=\chi _{ji}^{\ast }=|\chi _{ij}|e^{-ia_{ij}}$ and $a_{ij}$
is a phase field. Under the gauge transformation $a_{ij}\rightarrow
a_{ij}+\theta _{i}-\theta _{j}$, $f_{i}\rightarrow f_{i}e^{i\theta _{i}}$,
the effective action is invariant, because $\partial _{\tau }\theta _{i}$
term can be absorbed in the Lagrange multiplier. The gauge invariance thus
gives rise to a U(1) gauge field, which is also related to the local
constraint. By gauge fixing, $\chi _{ij}$ can be chosen to be real, and a
saddle point approximation leads to the $\pi $-flux state with
\begin{equation}
\chi _{i,i+x}=-\chi (-1)^{i_{y}},\text{ }\chi _{i,i+y}=\chi ,\text{ }
\label{PiFlux}
\end{equation}%
and the Lagrangian multiplier is fixed by the particle-hole symmetry at $%
\lambda _{i}=0$. Although the above mean-field ansatz breaks the lattice
translation symmetry, it can be restored when projecting the mean-field
state onto the physical Hilbert space with two spinons per site. Then the
effective model can be written as%
\begin{eqnarray}
H_{\pi \text{F}} &=&2JN\chi ^{2}-J\chi \sum_{i\alpha }(f_{i\alpha }^{\dag
}f_{i+y,\alpha }+f_{i\alpha }^{\dag }f_{i-y,\alpha }  \notag \\
&&-(-1)^{i_{y}}f_{i\alpha }^{\dag }f_{i-x,\alpha }-(-1)^{i_{y}}f_{i\alpha
}^{\dag }f_{i+x,\alpha }),
\end{eqnarray}%
which is quadratic in spinon operators and is easily diagonalized. The
spinon excitation spectrum is obtained by $\epsilon _{\mathbf{k}}^{\pm }=\pm
2J\chi \sqrt{\cos ^{2}k_{x}+\cos ^{2}k_{y}}$ with four gapless nodes at $%
k_{x}=\pm \frac{\pi }{2}$ and $k_{y}=\pm \frac{\pi }{2}$. The ground state
energy per site is determined by filling the negative spinon bands:%
\begin{equation}
\epsilon _{\pi \text{F}}=4\int_{-\pi }^{\pi }\frac{dk_{x}}{2\pi }\int_{-%
\frac{\pi }{2}}^{\frac{\pi }{2}}\frac{dk_{y}}{2\pi }\epsilon _{\mathbf{k}%
}^{-}+2J\chi ^{2},
\end{equation}%
where the minimum energy is attained at $\chi =0.958091$. When taking into
account the nodal SU(2) symmetry in the low energy limit, the $\pi $-flux
state gains an emergent SU(8) symmetry \cite{Hermele-2005}.

In fact, the low energy effective theory of the SU(4) $\pi $-flux states
should be described by a fluctuating massless U(1) gauge field minimally
coupled to the Dirac spinons. This is different from the $\pi $-flux state
with a gapless SU(2) gauge field in the spin-1/2 quantum Heisenberg
antiferromagnets. However, the SU(4) $\pi $-flux state resembles the SU(2)
staggered flux state. Recently, Hermele \textit{et al}. argued that if the
spin index is generalized to $N$, the problem of $2N$ two-component Dirac
fermions coupled to a compact U(1) gauge field is deconfined for sufficient
large $N$ \cite{Hermele-2004}. Furthermore, Assaad has performed quantum
Monte Carlo simulations of the SU($N$) antiferromagnetic Heisenberg model on
the square lattice and found that the optimal ground state for $N=4$ is a $%
\pi $-flux state with gapless spinon excitation and spin algebraic
correlations \cite{Assaad-2005}. The lattice translational and spin
rotational symmetries are preserved in such a $\pi $-flux state. Thus, a
system of interacting spin-$\frac{3}{2}$ fermionic ultracold atoms at
half-filling in a\ two-dimensional square optical lattice is a possible
experimental realization to observe this interesting algebraic spin liquid
state. The equal-time spin-quadrupole correlations and the SO(5) adjoint
operator correlation functions display the \textit{same} power-law behavior
due to the SU(4) symmetry%
\begin{equation}
(-1)^{(i_{x}+i_{y})}\left\langle n_{i}^{a}n_{0}^{a}\right\rangle
=(-1)^{(i_{x}+i_{y})}\left\langle L_{i}^{ab}L_{0}^{ab}\right\rangle \sim
\frac{1}{|\mathbf{r}_{i}|^{\eta }},
\end{equation}%
which in principle can be measured by spatial noise correlation experiments %
\cite{Altman-2004,Folling-2005}. In the mean field description, the
correlation functions in the $\pi $-flux state decay as $1/|\mathbf{r}|^{4}$%
, because the spinons are effectively described by (2+1) dimensional
relativistic free Dirac fermions at low energy regime. However, the gauge
fluctuations are expected to enhance the correlations functions as shown in
the quantum Monte Carlo results ($1/|\mathbf{r}|^{1.12}$) \cite{Assaad-2005}
and the calculations including the gauge fluctuations in a large-$N$
approximation \cite{Rantner-2001}.

\subsection{Effective model and the phase diagram}

According to the Landau's continuous phase transition theory, two phases
with incompatible symmetries can not be connected by a continuous phase
transition \cite{Senthil-2004}. In the previous section, our mean field
theory have shown that the quantum critical $\pi $-flux state separates two
insulating long-range ordered states, which are characterized by different
symmetries, order parameters, and topological characterizations. However, we
can not rule out the possibility that the quantum critical $\pi $-flux state
actually controls a region in the parameter space. Moreover, our saddle
point approaches to the generalized Hubbard model have neglected the
particle number fluctuations, however, such particle number fluctuations may
melt the assumed spin ordering at small but finite $c_{2}$ and stabilize the
algebraic spin liquid state. Our system provides such a chance to study the
quantum critical phase within the Landau's paradigm.

Let us start from the effective Hamiltonian in the strong coupling limit %
\cite{CWu-2006}%
\begin{equation}
H_{\text{eff}}=J_{1}\sum_{\langle ij\rangle
,a<b}L_{i}^{ab}L_{j}^{ab}+J_{2}\sum_{\langle ij\rangle
,a}n_{i}^{a}n_{j}^{a}+J_{3}\sum_{i,a<b}(L_{i}^{ab})^{2},
\end{equation}%
where $J_{3}\propto c_{2}$. At the SU(4) point, $J_{3}=0$ and $%
J_{1}=J_{2}=4t^{2}/c_{0}$, the effective model is reduced to the previous
SU(4) Heisenberg antiferromagnetic spin model. In the spin-quadrupole
ordering phase, we should have $J_{2}\gg J_{1}>0$ and $J_{3}>0$, the first
term can be neglected, and the effective model Hamiltonian becomes an SO(5)
quantum rotor model. However, in degenerate antiferromagnetic
spin-dipole/spin-octupole ordering phase, we should have $J_{1}\gg J_{2}>0$
and $J_{3}<0$, the latter two terms can be neglected, and the effective
model is reduced to an SO(5) Heisenberg antiferromagnetic spin model. Thus,
this effective model Hamiltonian well describes the basic physics of the
system from the strong coupling limit. In this effective model, the virtue
tunnelling processes of atoms between lattice sites have been taken into
account, and the various spin ordering formation can be regarded as possible
instabilities of algebraic spin liquid state.

\subsubsection{Staggered spin-quadrupole ordering state}

Since $J_{2}>J_{1}$ and $J_{3}>0$, we have to find relevant interactions for
instability of the $\pi $-flux state to spin-quadrupole fluctuations. Again
we use the fermionic spinon representations and the effective Hamiltonian
can be expressed as%
\begin{eqnarray}
H_{\text{eff}} &=&-J_{1}\sum_{\langle ij\rangle ,\alpha \beta }f_{i\alpha
}^{\dag }f_{j\alpha }f_{j\beta }^{\dag }f_{i\beta }+{}J_{-}\sum_{\langle
ij\rangle ,a}n_{i}^{a}n_{j}^{a}  \notag \\
&&-J_{3}\sum_{i,a}(n_{i}^{a})^{2},
\end{eqnarray}%
where $J_{-}=J_{2}-J_{1}$ and spinon number satisfy the constraint $%
\sum_{\alpha }f_{i\alpha }^{\dag }f_{i\alpha }=2$. Since $J_{-}>0$ and $%
J_{3}>0$, the first term is decoupled according to the valence bond order
parameters in the $\pi $-flux state as (\ref{PiFlux}) and the latter two
terms are treated in a mean field approximation. The following staggered
spin-quadrupole order parameter is introduced
\begin{equation}
(-1)^{i}\left\langle n_{i}^{4}\right\rangle =\Delta ,\text{ }%
(-1)^{i}\left\langle n_{i}^{a}\right\rangle =0\text{ }(a\neq 4)\text{.}
\end{equation}%
Then, the mean field Hamiltonian can be obtained as%
\begin{eqnarray}
H_{\text{MF}} &=&2J_{1}\chi \sum_{\mathbf{k},\alpha }\left[ \left( \cos
k_{x}\right) f_{\mathbf{k}\alpha }^{\dag }f_{\mathbf{k}-\pi \hat{k}%
_{y},\alpha }-\left( \cos k_{y}\right) f_{\mathbf{k}\alpha }^{\dag }f_{%
\mathbf{k}\alpha }\right]   \notag \\
&&-(2{}J_{-}+J_{3})\Delta \sum_{\mathbf{k},\alpha \beta }f_{\mathbf{k}\alpha
}^{\dag }\Gamma _{\alpha \beta }^{4}f_{\mathbf{k}-\mathbf{Q},\beta }  \notag
\\
&&+2J_{1}\chi ^{2}N+(2{}J_{-}+J_{3})\Delta ^{2}N,
\end{eqnarray}%
which can be easily diagonalized. The spinon quasiparticle spectrum is thus
derived as%
\begin{equation}
E_{\mathbf{k}}^{\pm }=\pm \sqrt{4J_{1}^{2}\chi ^{2}(\cos ^{2}k_{x}+\cos
^{2}k_{y})+(2{}J_{-}+J_{3})^{2}\Delta ^{2}},
\end{equation}%
where an energy gap is induced by the formation of the long range
spin-quadrupole ordering. By minimizing the ground state energy, a set of
self-consistent equations determining the order parameters $\chi $ and $%
\Delta $ are deduced to%
\begin{eqnarray}
\int_{-\frac{\pi }{2}}^{\frac{\pi }{2}}\frac{dk_{x}dk_{y}}{(2\pi )^{2}}\frac{%
\left( \cos ^{2}k_{x}+\cos ^{2}k_{y}\right) }{E_{\mathbf{k}}^{+}} &=&\frac{1%
}{8J_{1}},  \notag \\
\int_{-\frac{\pi }{2}}^{\frac{\pi }{2}}\frac{dk_{x}dk_{y}}{(2\pi )^{2}}\frac{%
4(2J_{-}+J_{3})}{E_{\mathbf{k}}^{+}} &=&1.
\end{eqnarray}%
From these equations, we can easily find the critical condition, i.e. the
boundary between the spin-quadrupole ordering and the $\pi $-flux state. By
setting $\Delta =0$, we find the phase boundary as%
\begin{equation}
\frac{J_{3}}{J_{2}}\approx -3.49\frac{J_{-}}{J_{2}}+1.49.
\end{equation}

\subsubsection{Degenerate antiferromagnetic spin-dipole/spin-octupole
ordering state}

Next we turn to the parameter regime in which spin-dipole and spin-octupole
fluctuations are dominant. To investigate the long-range order formation,
the effective Hamiltonian can be rewritten as
\begin{eqnarray}
H_{\text{eff}} &=&-J_{2}\sum_{\langle ij\rangle ,\alpha \beta }f_{i\alpha
}^{\dag }f_{j\alpha }f_{j\beta }^{\dag }f_{i\beta }-{}J_{-}\sum_{\langle
ij\rangle ,a<b}L_{i}^{ab}L_{j}^{ab}  \notag \\
&&+J_{3}\sum_{i,a<b}(L_{i}^{ab})^{2}.
\end{eqnarray}%
Since $J_{-}<0$ and $J_{3}<0$, the valence bond order parameter as the $\pi $%
-flux state as (\ref{PiFlux}) can be chosen for the first term and a mean
field approximation can be made for the last two terms. Then the degenerate
antiferromagnetic spin-dipole/spin-octupole order parameter is introduced by%
\begin{equation}
(-1)^{i}\left\langle L_{i}^{15}\right\rangle =\Delta ,\text{ }%
(-1)^{i}\left\langle L_{i}^{ab}\right\rangle =0\text{ }(\text{otherwise})%
\text{.}
\end{equation}%
Thus, the mean field Hamiltonian is obtained as%
\begin{eqnarray}
H_{\text{MF}} &=&2J_{2}\chi \sum_{\mathbf{k},\alpha }\left[ (\cos k_{x})f_{%
\mathbf{k}\alpha }^{\dag }f_{\mathbf{k}-\pi \hat{k}_{y},\alpha }-(\cos
k_{y})f_{\mathbf{k}\alpha }^{\dag }f_{\mathbf{k}\alpha }\right]   \notag \\
&&-(J_{3}+2{}J_{-})\Delta \sum_{\mathbf{k},\alpha \beta }f_{\mathbf{k}\alpha
}^{\dag }\Gamma _{\alpha \beta }^{15}f_{\mathbf{k}-\mathbf{Q},\beta }  \notag
\\
&&+2J_{2}\chi ^{2}N-(J_{3}+2{}J_{-})\Delta ^{2}N,
\end{eqnarray}%
which yields the spinon quasiparticle spectrum%
\begin{equation}
E_{\mathbf{k}}^{\pm }=\pm \sqrt{4J_{2}^{2}\chi ^{2}(\cos ^{2}k_{x}+\cos
^{2}k_{y})+(J_{3}+2{}J_{-})^{2}\Delta ^{2}},
\end{equation}%
where again an energy gap is created due to the formation of the long-range
spin ordering. The corresponding mean field self-consistent equations are
derived as%
\begin{eqnarray}
\int_{-\frac{\pi }{2}}^{\frac{\pi }{2}}\frac{dk_{x}dk_{y}}{(2\pi )^{2}}\frac{%
\cos ^{2}k_{x}+\cos ^{2}k_{y}}{E_{\mathbf{k}}^{+}} &=&\frac{1}{8J_{2}},
\notag \\
\int_{-\frac{\pi }{2}}^{\frac{\pi }{2}}\frac{dk_{x}dk_{y}}{(2\pi )^{2}}\frac{%
4(J_{3}+2{}J_{-})}{E_{\mathbf{k}}^{+}} &=&-1.
\end{eqnarray}%
Assuming $\Delta =0$, the phase boundary between the $\pi $-flux phase and
the degenerate antiferromagnetic spin-dipole /spin-octupole ordering phase
can be found as%
\begin{equation}
\frac{J_{3}}{J_{2}}\approx -2\frac{J_{-}}{J_{2}}-1.49.
\end{equation}

\begin{figure}[tbp]
\includegraphics[scale=0.5]{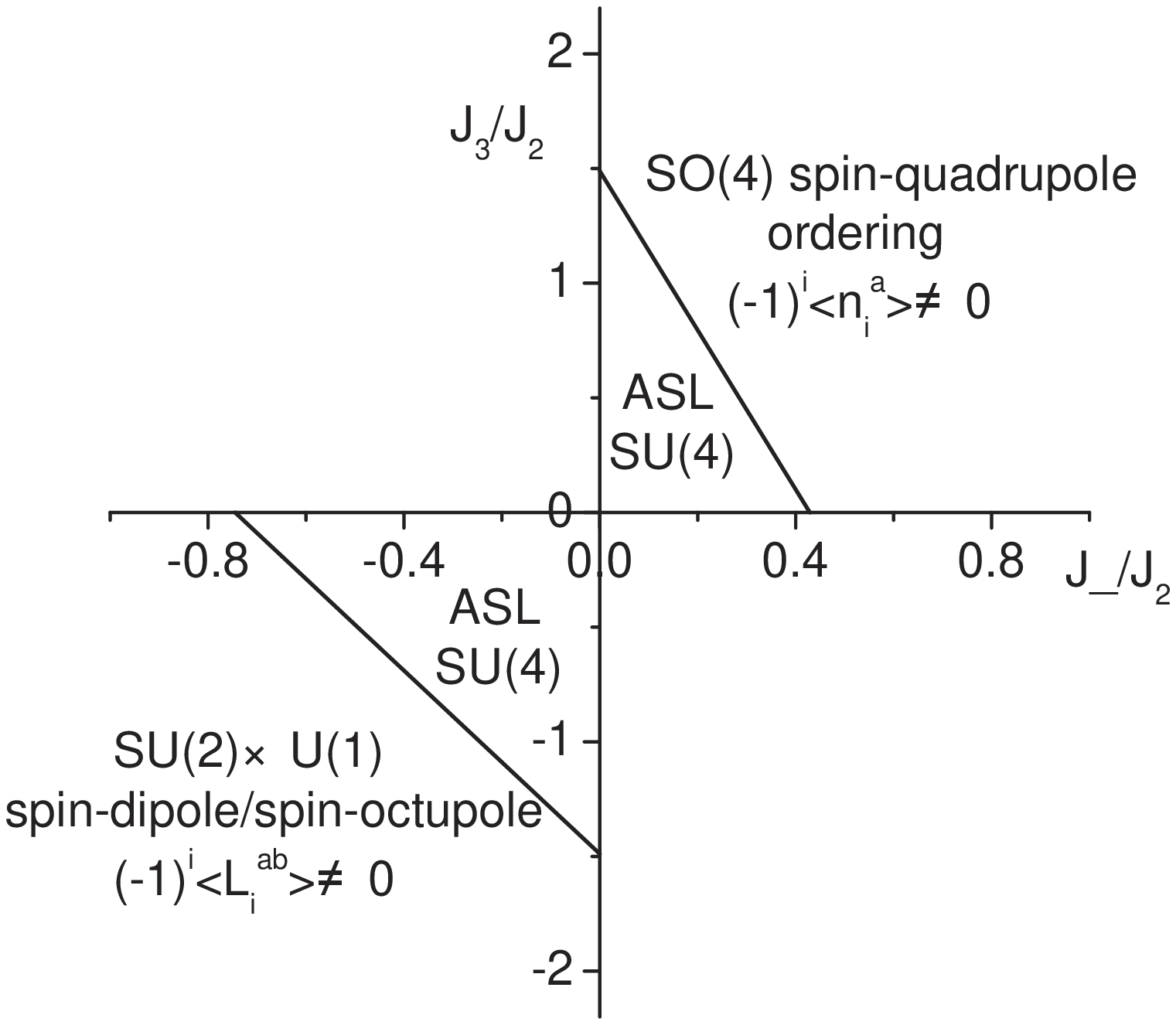}
\caption{The phase diagram for the effective model in the strong coupling
limit. ASL stands for algebraic spin liquid and $J_{-}=J_2-J_1$.}
\end{figure}

Now let us map out a ground state phase diagram of the model, which is shown
in Fig. 4. When $c_{0}\gg |c_{2}|$ and $c_{0}\gg t$, we have $|J_{-}|\sim
t^{2}|c_{2}|/c_{0}^{2}\ll |c_{2}|$ and $J_{3}\sim c_{2}$. Thus a stable $\pi
$-flux state with spin algebraic correlations may be observed in the region
that $|c_{2}|$ is comparable to $t^{2}/c_{0}$, implying that the competition
of the exchange interaction between different sites and the on-site spin
dependent interaction determines the low energy physics in the critical
region. Furthermore, two second-order phase transitions occur along the two
phase boundary curves, respectively. We would like to emphasize that there
is no direct phase transition between the spin-quadrupole ordering and
spin-dipole/spin-octupole ordering states.

\section{Discussion}

We have considered Mott insulating phases of spin-$\frac{3}{2}$ fermionic
atoms in two-dimensional square optical lattices at half filling. In
addition to the spin-quadrupole ordering phase, we found a degenerate
antiferromagnetic spin-dipole/spin-octupole ordering phase. This is a
symmetry breaking phase with the Goldstone manifold SO(5)/[SU(2)$\times $%
U(1)]=CP$^{3}$ and six Goldstone bosonic modes. Compared to spin-quadrupole
ordering phase, the time reversal symmetry is broken, and there are coupled
transverse bosonic modes, which result in a suppression of the density wave
velocity in the strong coupling limit. The topological properties in the two
symmetry breaking phases are quite different: the spin-quadrupole ordering
state contains a non-abelian SU(2) Berry phase and is characterized by the
second Chern number, while the degenerate spin-dipole/spin-octupole ordering
state contains only abelian U(1) Berry phase and is described by the first
Chern number.

Due to these fundamental differences, a continuous phase transition from one
to another is almost impossible (although a first order transition cannot be
excluded in principle). Between these two phases, the quantum fluctuations
are so strong that an SU(4) $\pi $-flux spin liquid state may be regarded as
a quantum critical phase. In the low energy excitations of the algebraic
liquid phase, there emerge U(1) massless gauge field and fractionalized spin-%
$\frac{3}{2}$ gapless spinons. Both the equal-time correlation functions of
the spin-quadrupole operators and the SO(5) adjoint operators display the
same power law behavior at long distance. We expect these features can be
detected by spatial noise correlation experiments in future. We should
mention that the above results were obtained in the mean field theory and
more precise treatment of quantum fluctuations may modify the scenario.

\begin{acknowledgments}
The authors are grateful to Congjun Wu and Xiao-Gang Wen for their
stimulating discussions. We acknowledge the support of NSF-China (Grant
No.10125418 and Grant No.10474051) and the national program for basic
research.
\end{acknowledgments}

\end{document}